\documentclass[sigconf]{acmart}
\AtBeginDocument{%
  }

\setcopyright{acmlicensed}
\copyrightyear{2018}
\acmYear{2018}
\acmDOI{XXXXXXX.XXXXXXX}
\acmConference[Conference acronym 'XX]{Make sure to enter the correct
  conference title from your rights confirmation email}{June 03--05,
  2018}{Woodstock, NY}
\acmISBN{978-1-4503-XXXX-X/2018/06}

\begin{document}

\title{Crisis Messaging Journeys: How Past Guidance Undermines Present Authority}

\author{Tawfiq Ammari}
\email{tawfiq.ammari@rutgers.edu}
\affiliation{%
  \institution{Rutgers School of Communication and Information}
  \city{New Brunswick}
  \state{NJ}
  \country{USA}
}

\renewcommand{\shortauthors}{Ammari}

\begin{abstract}
This study examines how the U.S. Centers for Disease Control and Prevention (CDC) communicated COVID-19 guidance on Twitter and how publics responded across two years of the pandemic. Drawing on 275,124 tweets mentioning or addressing @CDCgov, I combine BERTopic modeling, sentiment analysis (VADER), credibility checks (Iffy Index), change-point detection (PELT), and survival analysis to map three discourse phases: (1) early hoax and testing debates, (2) lockdown and mask controversies, and (3) post-vaccine variant concerns. I introduce the concept of crisis messaging journeys to explain how archived “receipts” of earlier CDC statements fueled epistemic struggles, political polarization, and long-term engagement. Findings show that skeptical, cognitively complex discourse—especially questioning institutional trust—sustained participation, while positive affirmation predicted quicker disengagement. I conclude with design recommendations for annotated, cautious, and flashpoint-responsive communication to bolster trust and resilience during protracted public health crises.
\end{abstract}



\received{20 February 2007}
\received[revised]{12 March 2009}
\received[accepted]{5 June 2009}

\maketitle

\section{Introduction}
Public health crises are marked by uncertainty, confusion, and heightened emotion, which fuel public discourse—particularly on social media. This discourse typically centers on two key concerns: the risk of infection and the consequences of crisis management measures \citep{gui_et_al_18}. As crises unfold, both public responses and official messaging evolve, especially during flashpoints such as the first local case or the infection of a health worker \citep{coombs2021ongoing}. These critical moments intensify attention and demand rapid shifts in communication strategy \citep{dalrymple2016facts}. Yet, effectively conveying public health information remains a major challenge, as illustrated by the U.S. measles resurgence, which underscored persistent problems such as vaccine misinformation, healthcare access disparities, and low health literacy. These challenges underscore the importance of communication channel design in shaping how crisis organizations like the CDC use social media to disseminate information and engage the public. As the temporal distance from the initial crisis event grows, adapting messages becomes even more crucial \citep{kim2020organizational,kim2022time}. This raises pressing questions for crisis communication scholars: How do the affordances and limitations of social media platforms \citep{treem2013social} influence the reach, credibility, and interactivity of evolving public health messages? To understand evolving public health messaging, especially at important junctures in time, I ask:

\textbf{RQ1: What are the COVID flashpoints in publics responding to the CDC?}

Twitter operates as a virtual “public square” where ad hoc publics coalesce around hashtags that capture the attention of interested audiences \citep{jenzen2021symbol, bruns_structural_2014, livingstone2005relation}. These publics also take shape through user mentions, which are both searchable and visible to others \citep{kountouri2023polarizing}. Platform affordances—such as retweets, quote tweets, and replies—play a key role in how frames circulate and are negotiated within networked publics \citep{faraj_materiality_2013, leonardi_social_2017}.

While social media can deliver timely public health information, it also facilitates the spread of conspiracy theories, often fueled by fear, uncertainty, and lack of trust in institutions like the CDC \cite{jones_et_al_24}. During the Zika outbreak, for example, social media served as both a news source and a conduit for rumors and conspiracy, partly due to shortcomings in official communication \citep{kou_at_al_17, gui_et_al_17}. Misleading content circulated widely through rich media (e.g., memes and images) often full of hate and discriminatory messaging \citep{wang_et_al_23, liao_23, Vishwamitra_24, javed_et_al_23}. Traditionally scientific ways of communicating medical and scientific information through visualizations were also adopted to spread minisformation by misreading or misrepresenting visual information \citep{lisnic_et_al_24,lee_et_al_21}.

The COVID-19 pandemic saw the rise of medical and scientific populism, with figures like Dr. Judy Mikovitz—herself a health professional—challenging consensus on masking and vaccination \citep{premat2024introduction}. Verified accounts can also become “perceived experts” who spread vaccine misinformation \citep{harris2024perceived}, particularly concerning the side effects of vaccines like Oxford/AstraZeneca \citep{hobbs2024low}. As Bruns observes, Twitter communities develop shared norms and structures through ongoing, collaborative participation, with long-term contributors shaping the platform’s cultural fabric over time \citep{bruns2008produsage}. Given the importance of veteran and involved users, I ask:

\textbf{RQ2: What factors contribute to a user's long-term influence in these spaces, and how do they shape messenger impact on public health publics?}

To address these questions, I analyze longitudinal discourse patterns on Twitter directed at the CDC during the COVID-19 pandemic. By identifying key flashpoints and examining user engagement over an extended crisis timeline, I aim to contribute to both crisis informatics and CSCW research. Our study highlights how publics evolve in their reactions to changing public health messaging and how particular user characteristics sustain long-term participation in health-related discussions. In doing so, I offer insights into designing more adaptive, resilient communication strategies for public health organizations navigating protracted crises.

Building on theories of data journeys \citep{edwards2013vast} and political epistemology \citep{Freelon03032020}, this study examines how crisis messages are not merely transmitted but are transformed across time, space, and sociopolitical contexts. I introduce the concept of \textit{crisis messaging journeys} to capture how initial CDC messages, once disseminated, are subject to reinterpretation, resistance, and repurposing in ways that shape the public understanding of science itself. Through a mixed-methods approach combining change-point detection and survival analysis, I trace how public discourse around CDC guidance evolved, identifying key flashpoints that destabilized trust and reconfigured the epistemic terrain of pandemic communication.

By situating public health communication within broader struggles over political knowledge production, this work challenges conventional, linear models of information dissemination. In particular, it moves beyond the deficit model of misinformation—which assumes that misinformation thrives primarily due to a lack of accurate knowledge \citep{simis2016lure}—toward a perspective that foregrounds the active role of digital publics in shaping what counts as knowledge. Digital publics, far from being passive recipients, engage in the construction of alternative epistemologies around risk, uncertainty, and authority, often contesting institutional narratives in the process \citep{MilanVelden2016}.

\section{Related Work}
This section situates our study within existing scholarship on crisis communication, epistemic trust, and the temporal dynamics of online discourse.

First, we review research on \emph{epistemic governance and public health messaging}, which frames truth and credibility as politically contested and highlights how shifting guidance can destabilize institutional trust during fast-moving crises \cite{jones_chand_24}. 

Next, we turn to the \emph{Crisis and Emergency Risk Communication (CERC) literature}, emphasizing its call for two-way communication and exploring why traditional top-down approaches---like those often used by the CDC---struggle to adapt over long, unpredictable crisis timelines \cite{reynolds2005crisis}. 

Finally, we draw on studies of \emph{flashpoints and temporal change} in health crises to explain how sudden events, such as new variants or policy reversals, create discontinuities in public sentiment and knowledge production. Together, these strands of work motivate our focus on ``crisis messaging journeys'' as a way to understand how CDC guidance travels, mutates, and encounters resistance over time.

\subsection{Contesting Truth Online: Epistemic Trust and Public Health Messaging During COVID-19} \label{related_work_1}
This approach aligns with the concept of epistemic governance introduced by \citet{Alasuutari02012014}, which emphasizes how contemporary governance increasingly operates through the management of meaning. By influencing actors’ understandings of reality, their identities, and normative goals, policymakers strategically shape public perceptions to advance particular agendas. In this context, truth itself becomes a site of contestation and control for trust, testimonial credibility, and legitimacy risks \citep{jones_chand_24}.

For institutions like the CDC to be stable, they need to justify themselves to the public by evading legitimacy shocks to their epistemic trust \citep{tucker2019unelected,tucker2024global} in the organization \cite[P.5]{jones_chand_24}. Fuerstein \cite{fuerstein2013epistemic} contends that justification—meeting specific epistemic standards—is a necessary precondition for warranted epistemic trust \cite[P.5]{jones_chand_24}. He likens political epistemology to a prisoner’s dilemma in which skepticism is rational, and sees justification as the “cooperation technology” that makes collective trust possible. Epistemic trust and legitimacy, on this account, are therefore tightly linked. Building on this, Jianing \cite{jianing_25} complicates the picture of skepticism in digital spaces, arguing it is not monolithic. She distinguishes accuracy-motivated skepticism, which can foster critical engagement and learning, from identity-motivated skepticism, which often entrenches misinformation by reinforcing group loyalties and emotional attachments. This differentiation is crucial for understanding how misinformation persists in polarized or ideologically charged environments.

Institutions whose authority depends on legitimacy or that are exposed to the risks of misinformation need to pay close attention to how epistemic trust and distrust circulate within the communities they serve. This is a broad imperative, encompassing nearly all governmental bodies and many corporate actors. Public health agencies, for example, have already witnessed how misinformation can undermine vaccination campaigns \cite{tanzer2024role}. The Centers for Disease Control and Prevention (CDC) exemplify such reliance on epistemic trust—the willingness to treat knowledge communicated by the CDC as ``significant, personally relevant, and transferable across contexts.'' \cite{tanzer2024role}. In other words, the CDC’s effectiveness hinges on maintaining strong testimonial credibility. Systematically tracking the dynamics of (dis)trust, not merely the spread of misinformation, is therefore essential for the functioning and legitimacy of these institutions \cite{jones_chand_24}.

The COVID-19 pandemic exposed the dynamic and often contentious relationship between institutional public health messaging and public reception. As new scientific evidence emerged, guidelines issued by agencies like the CDC evolved—sometimes dramatically—prompting both confusion and backlash from digital publics. Social media platforms, particularly Twitter, served as critical arenas where these evolving guidelines were contested, negotiated, and reframed \citep{plantin_et_al_2018}.

One major point of contention was the CDC’s and World Health Organization’s (WHO) reluctance to acknowledge early findings on the airborne transmission of COVID-19 \citep{albarracin2024health,early_masking_2023,mask_early_2}. Early in the pandemic, the CDC advised against mask use. As new evidence accumulated, it reversed course in April 2020 to endorse masking as essential. Later, the agency linked mask guidance to fluctuating community transmission levels. These shifting rationales and criteria produced a fragmented and often confusing message over time and across regions \citep{albarracin2024health}.

At the same time, the urgency to share findings quickly fueled a surge in the use of preprint servers such as MedRxiv, especially in the pandemic’s early months \citep{lachapelle2020covid,lachapelle2022scientific}. This rapid dissemination sometimes led to the circulation of unvetted studies that were later retracted, yet widely cited and shared on social media. A notable example is the early study claiming the efficacy of hydroxychloroquine against COVID-19—a result subsequently withdrawn \cite[pp.28–30]{fraser2021evolving}. Managing redactions became an important problem to solve for health scientists in general, and preprint servers in particular. Clear communication about research that has been retracted became a key principle in engaging the public \citep{fraser2021evolving}. One element of this approach is illustrated as follows:
“In cases where a journal retracts or fully removes an article that also exists as a preprint, the editors should inform the preprint server so that it can determine whether to issue a withdrawal notice, provide a retraction link or notice, or take another action such as releasing a new version or removing the content” \cite{beck2020building}.

Taken together, these insights illuminate a shifting communicative landscape where public health communication is entangled in broader epistemic and political dynamics. It must therefore grapple not only with the clarity and accuracy of messages but also with the diverse, often conflicting epistemic orientations of audiences. This becomes particularly salient in moments of high uncertainty—such as during outbreaks of emerging infectious diseases—where institutional trust is volatile and the legitimacy of expertise is frequently contested.

\subsection{Crisis and Emergency Risk Communication}
When crises such as pandemics or natural disasters disrupt daily life, people face heightened demands for rapid, reliable information and mutual aid. Online forums and social media capture and amplify this urgency, serving as spaces where collective resilience takes shape \citep{stephens2022social, iyengar2024resilience}. Digital tools that help individuals manage acute pressures—whether sudden flooding or spikes in infection rates—become essential supports for strengthening community resilience \citep{iyengar2024resilience}.

COVID-19 made these patterns concrete. Digital platforms rapidly enabled the formation and coordination of mutual-aid networks that addressed immediate needs—food insecurity, job loss, and barriers to healthcare—through actions that spanned online spaces and neighborhood efforts; many later expanded to longer-term structural concerns such as housing affordability and racial justice \citep{10.1145/3490632.3490666, Haesler_et_al_stronger_together_21, soden_et_al_21}. At the same time, social distancing policies were accompanied by a marked rise in social media activity \citep{cho2023bright}, as people turned to these channels to sustain relationships, seek reliable information, and exchange emotional and material support—particularly among those ill with COVID-19 \citep{Jangid_et_al_together_alone_24}.

The Crisis and Emergency Risk Communication (CERC) model describes how, to better manage uncertainty at the height of crisis, an organization has to engage in two-way communication \citep{reynolds2005crisis} where the public can give feedback about the management of a crisis \citep{renn2015stakeholder}. Specifically, public facing organizations like the CDC or FEMA need to engage in what \cite{waters2011tweet} call symetric two-way communication which provides for a mutual understanding between the public and organizations which could lead to changes in their communication strategies \citep{grunig2008excellence,waters2011tweet}. However, earlier work has shown that CDC's online communication has been mostly top-down (from the CDC to users) instead of bottom-up communication (from users to the CDC; \citep{gesser2014risk}). 

One major challenge faced by scholars lies in the lack of a mechanism to determine different phases of an outbreak (beginning, middle, and end) since they are not predictable, especially in IEDs \cite[P.457]{dalrymple2016facts}. Zhang et al. \citep{zhang2019social} argue that time is one of the dimensions that needs to be addressed and understood for an intelligent crisis information system. After an initial crisis event—such as the emergence of a novel virus—public health organizations focus on communicating disease detection (e.g., symptoms) and prevention (e.g., protective measures) \cite{umphrey2003effects,higgins2006health}. In the maintenance phase of the crisis, public health organizations need to communicate healthcare services (services provided by healthcare system), scientific advances like discovering new evidence about the disease, and lifestyle risk factors (e.g., personal habits associated with the disease) \cite{ngai2020grappling}. However, CERC does not provide explicit guidelines for evolving health communication needs during ``a global maintenace phase that is months in duration'' \citep[P.5]{miller2021being} much like COVID.  

\subsection{Flashpoints of change in a health crisis} \label{related_work:flashpoint}
Earlier research illustrates that crisis communication is temporally dynamic. As more time elapses from the peak of a crisis, public sentiment on social media often grows more positive and institutions regain support—observed in contexts involving FEMA and pharmaceutical firms \citep{kim2020organizational, kim2022time}. However, during COVID-19, this pattern did not hold. While Twitter usage spiked during lockdowns and early containment efforts in March 2020, sentiment sharply declined as the pandemic persisted \citep{valdez2020social}, underscoring the limits of institutional messaging and the shifting emotional terrain of crisis publics over time.

Events such as the first transmission of a disease from a patient to a healthcare worker—for example, Ebola spreading from a patient to a nurse—serve as “flashpoints of change” \citep{vanderford2007emergency}, triggering a need for “rapid redirection on social media” \citep{dalrymple2016facts}. In such moments, both the public and institutions are confronted with the precarious balance between certainty and uncertainty. As \citet{fox2012medical} argues, medicine inherently involves embracing “inevitable uncertainty,” where scientific knowledge is provisional—what he calls “certainty for now.” This epistemic fragility is exacerbated in the context of emerging infectious diseases (EIDs) like Ebola or COVID-19, where public health authorities often rely on existing protocols that create an illusion of certainty \citep{dalrymple2016facts}. For instance, when the CDC’s Ebola protocols failed to prevent transmission, their response focused on reinforcing what was already known about the disease rather than acknowledging the breakdown in assumptions \citep[P.455]{dalrymple2016facts}.

When institutions like the CDC present an overly confident narrative during chaotic periods, it may inadvertently amplify public uncertainty and erode institutional trust \citep{dalrymple2016facts}. In this context, the dynamics of crisis communication are deeply shaped by the medium through which information circulates. Twitter’s affordances—persistence, visibility, association, and scalability—significantly shape how information is shared and received during health emergencies \citep{treem2013social}. While these features enable broad and rapid dissemination, they also present complications: the persistence of outdated or misleading messages can hinder efforts to communicate timely updates, especially as situations evolve rapidly \citep{vaast_2017}.


Just as technologies and community networks are vital in managing acute crises, it is equally important to support communities grappling with chronic, recurring stressors such as poverty, health disparities, and systemic injustice \citep{iyengar2024resilience}. Social media platforms, far from being transient crisis tools, function as enduring infrastructures for collective resilience, community building, and socio-political transformation.

\section{Methods}
This section details the \emph{mixed-methods approach} we designed to capture the evolving life of CDC-related discourse on Twitter.

I begin by describing \emph{data collection and scope}, which spans 275{,}124 original tweets (plus retweet counts) from January 2020 to January 2022, filtered for U.S.-based, English-language posts mentioning or addressing the CDC.

I then explain the \emph{computational analyses} used to trace discourse shifts: BERTopic modeling to identify 71 topics grouped into five thematic clusters; VADER sentiment scoring; credibility checks via the Iffy Index; and extraction of platform features such as retweet counts, quote tweets, and rich-media usage.

Next, I outline \emph{temporal and survival analyses}, including Pruned Exact Linear Time (PELT) change-point detection to locate flashpoints in sentiment and topic prevalence, log-likelihood ratio tests to characterize linguistic change, and a Cox proportional hazards model to understand how linguistic style, sentiment, and topic focus influenced the longevity of user engagement.

Finally, I highlight \emph{ethical safeguards}, such as paraphrasing rather than quoting individual tweets and focusing on public institutions, ensuring that our computational methods remain grounded in responsible qualitative interpretation.

\subsection{Data Collection}
I used the twarc Python library for data mining via the Twitter API to collect the tweets used in this study. This was possible through the use of an individualized bearer token granted by Twitter. The inclusion criteria for the tweets in this study were tweets that were produced by users in the United States, in English, posted between January 2020 and January 2022, and not retweets. Furthermore, I specifically only included tweets that were sent to the official @CDCgov account, from @CDCgov, used the hashtag ‘\#CDC’, tagged @CDCgov, or mentioned the string ‘CDC’. In total, I collected 21,843 tweets sent to @CDCgov, 11,810 from @CDCgov, 20,012 using ‘\#CDC’, 53,104 tagging @CDCgov, and 162,295 mentioning the string ‘CDC’. Finally, I collected any other Tweets using the same ``Conversation ID'' which indicate that they are under the same discussion thread. In total, 275,124 tweets were collected. While I did not collect all retweets, I do have the number of times each tweet is retweeted with a total of 3,331,350 retweets. 

\subsection{Features Used in Temporal Analysis}
\subsubsection{Topic Modeling and Qualitative Analysis of Discourse Themes}
I applied BERTopic \citep{grootendorst2022bertopic}, a contextualized topic modeling technique that uses BERT embeddings clustered via HDBSCAN, to identify 71 coherent topics. Topic coherence is a critical consideration in evaluating topic model quality, as coherence scores have been shown to strongly correlate with human judgments of topic interpretability \citep{roder_exploring_2015}. To optimize model selection, I trained 25 BERTopic models with varying minimum cluster size parameters and selected the model with the highest coherence score (0.39) as the final model for analysis.

Following best practices in crisis informatics and theme detection \citep{vieweg_et_al_2010},  I was one of two coders who independently sampled and qualitatively analyzed tweets from each topic, iteratively describing, coding, and refining topic labels. Weekly consensus meetings were held to reconcile differences and converge on final topic interpretations—an essential step to ensure reliability and validity in computational discourse analysis \citep{McDonald_et_al_19}. We then grouped the interpreted topics into five overarching discourse themes that captured the major areas of discussion, including mitigation measures, vaccines, political discourse, and misinformation.

In addition to ensuring thematic validity, we also considered ethical standards for the use of publicly available social media data. \cite{fiesler_participant_2018} argue that Twitter users do not expect to be quoted verbatim in academic research, even if they are not named. They recommend adopting \cite{bruckman_studying_2002}'s levels of user disguise to protect participants' privacy. Following this guidance, I primarily described the content of tweets without direct quotations. I only named public organizations such as the CDC and news outlets when necessary.

\begin{figure*}
    \centering
    \includegraphics[width=\textwidth]{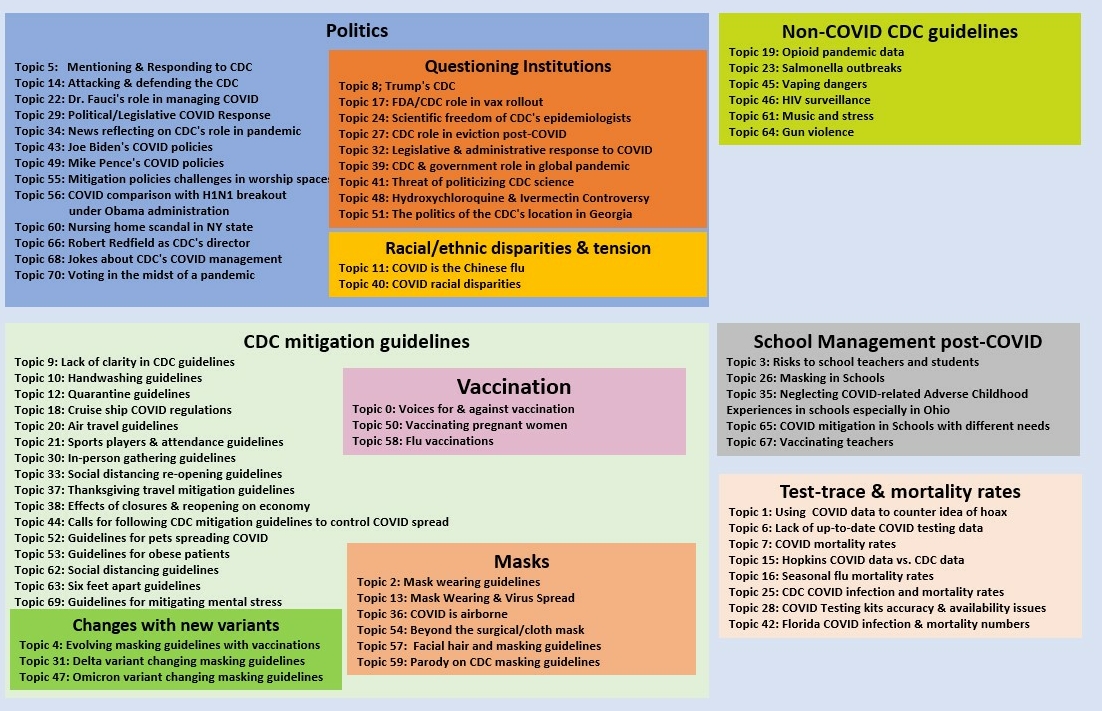} 
    \caption{Seventy-one CDC-related Twitter topics organized into five overarching themes—Politics, Non-COVID CDC guidelines, CDC mitigation guidelines, Vaccination, and Masks—plus two cross-cutting areas (School Management post-COVID and Test-trace and mortality rates). The figure highlights how discussions ranged from institutional trust and racial/ethnic tensions to evolving pandemic policies}
    \label{fig:top_themes}
\end{figure*}

Figure \ref{fig:top_themes} shows the breakdown of the 71 identified topics into five major themes reflecting the public discourse around COVID-19 mitigation and political debates.

\subsubsection{Sentiment Analysis} 
Tweet sentiment was calculated using the VADER tool \citep{hutto2014vader}, with composite scores which I used to label tweets as positive, negative, and neutral.

\subsubsection{Credibility Analysis.} 
I flagged tweets linking to low-credibility news sources using the Iffy Index \citep{pierri_2023}, which classifies domains based on factual reporting and ideological bias.

\subsubsection{Platform Signals.} 
I extracted features including retweet count, reply count, like count, quote count, rich media usage (e.g., images or videos), and verification status to understand patterns of information propagation and audience reach.

\subsection{Temporal Analysis}
To detect significant shifts in public discourse, I used the Pruned Exact Linear Time (PELT) change-point detection algorithm \citep{Truong2020}. PELT identifies points in a time series where the mean or variance significantly changes, allowing for precise localization of temporal discontinuities. I focused on shifts in both sentiment and topic prevalence and verified trend directions using the Mann-Kendall test \citep{hussain2019pymannkendall}, and characterized pre- and post-flashpoint discourse through log-likelihood ratio (LLR) analysis.

LLR determines whether particular terms have “topic signatures” that statistically define the “aboutness” \citep{gupta_07} of one set of texts compared to another. By comparing word usage distributions before and after detected change points, LLR identifies the most distinctive shifts in language framing. This method has been previously applied by  \cite{chancelor_et_al_18} to surface normative and linguistic differences between online communities, highlighting its suitability for tracing evolving public discourse in dynamic settings like health crises.

\subsection{Survival Analysis}

To model user engagement over time, I conducted survival analysis following the method presented in \cite{gao_et_al_21}. User commitment was defined as the duration of participation in CDC-related discourse in months. Predictors included platform signals (e.g., retweet count, rich media use, verified status) and engagement with the five main discourse themes. Users were considered censored if active at the time of data collection, and inactive if silent for three months. A Cox proportional hazards model (concordance = 0.612) revealed that rich media usage, verification status, and posting low-credibility sources were associated with sustained engagement.

\section{Tweeting through the Pandemic: From Hoaxes to New Variants} \label{sec:time}
In my findings, I show that there are three main phases: 
(1) hoaxes and early pandemic effects; (2) lock-downs and masks; and (3) post-vaccine variants and evolving guidelines.

\subsection{Beginning Phase: Hoaxes and Early Pandemic Effects}
\begin{figure*}
    \centering
    \includegraphics[width=\textwidth]{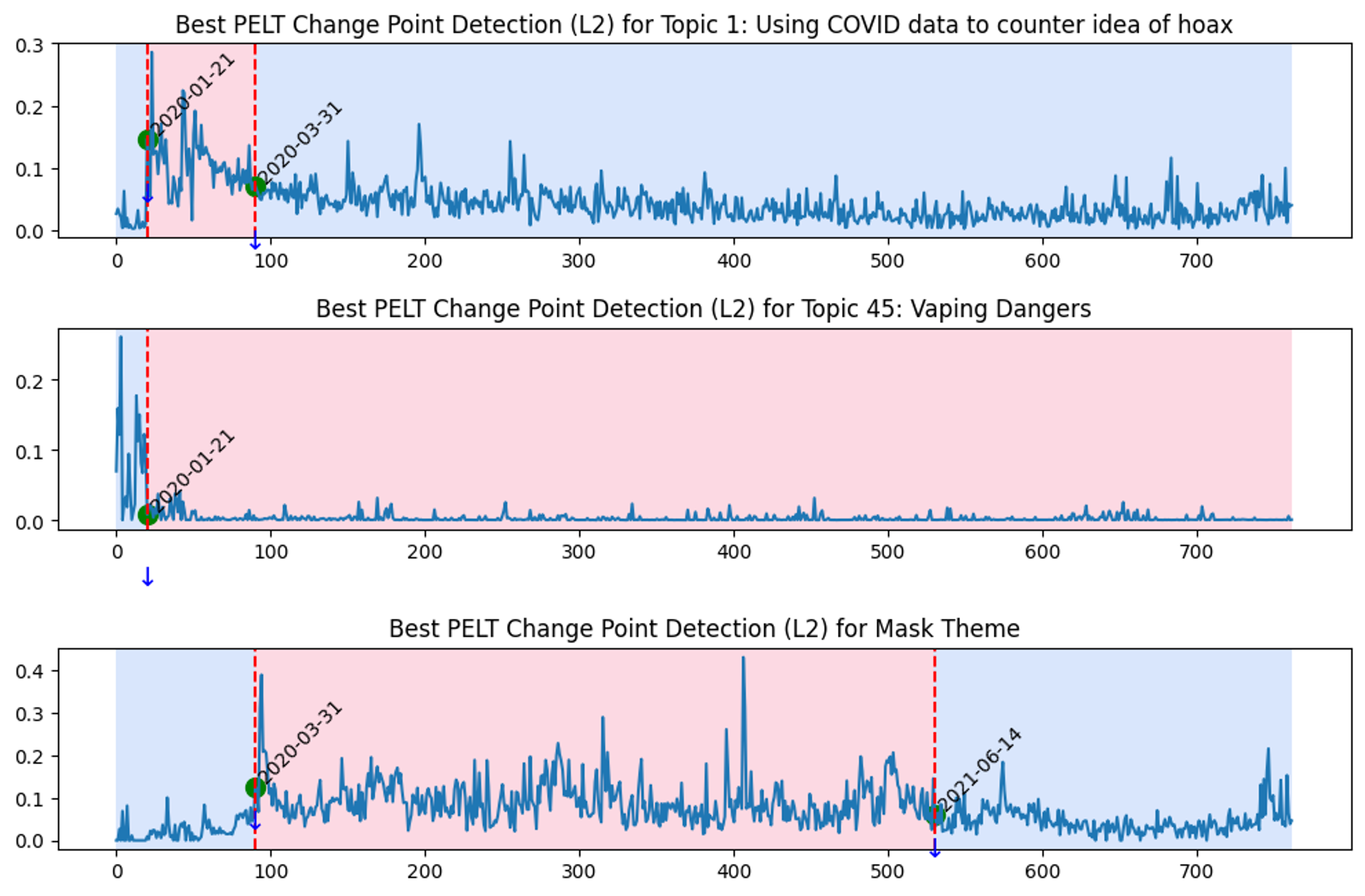}
    \caption{Change-point detection of three representative discourse streams using the PELT algorithm. Panels show (top) Topic 1 (‘COVID hoax’), (middle) Topic 45 (vaping dangers), and (bottom) the mask theme. Pink-shaded periods mark intervals of elevated change activity, when topic prevalence was especially volatile. Green dots denote statistically significant change-points and blue arrows indicate the direction of trend shifts before and after each change, highlighting how attention to these issues intensified or waned over time.}
    \label{fig:PELT_topics}
\end{figure*}

Figure \ref{fig:PELT_topics} shows that in the early days of 2020, COVID-19 started to have the definition of an acute emergency, with LLR values showing increased prevalence for Topics 62 (social distancing guidelines), 59 (parody of CDC guidelines), 57 (facial hair and mask guidelines), and 34 (news reflecting the CDC's role in managing the pandemic) after the first change-point.

\begin{figure*}
    \centering
    \includegraphics[width=\textwidth]{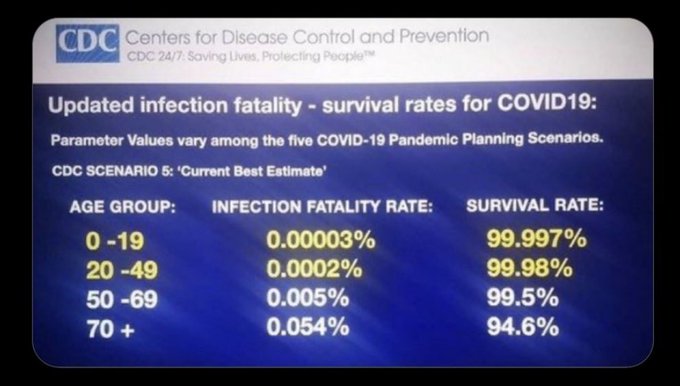}
    \caption{Example of a widely shared ‘receipt’—a screenshot of CDC infection-fatality data—used to argue that the agency overstated the severity of COVID-19. Although the underlying numbers are accurate, the image was reframed on Twitter to question the CDC’s interpretation and label the pandemic a hoax}
    \label{fig:hoax}
\end{figure*}

Subsequently, Topic 1 (using COVID data to counter the idea of a hoax) showed two change-points, on January 21\textsuperscript{st} and January 31\textsuperscript{st}, 2020. Both trends were downward, indicating this topic peaked between those dates. Both change-points for Topic 1 were accompanied by surges in discussions of CDC guidelines (Topics 62, 59, 57, and 34), highlighting growing public attention to the CDC's messaging as the crisis intensified. While some Tweets urged seriousness, others used satire or skepticism. Some of the ``receipts''  were used to question the severity of COVID and argue that the focus on its effects should death rates as opposed to infection rates, arguing that COVID did not have as much of an effect as the CDC was suggesting, in turn referring to CDC messaging as a hoax (an example of a receipt is presented in Figure \ref{fig:hoax}. It is important to note that users were not engaging directly in misinformation, but were questioning the meaning CDC messaging assigned to the same statistics. In effect, they were questioning the severity of the pandemic given the numbers provided by the CDC. 

\begin{figure*}
    \centering
    \includegraphics[width=\textwidth]{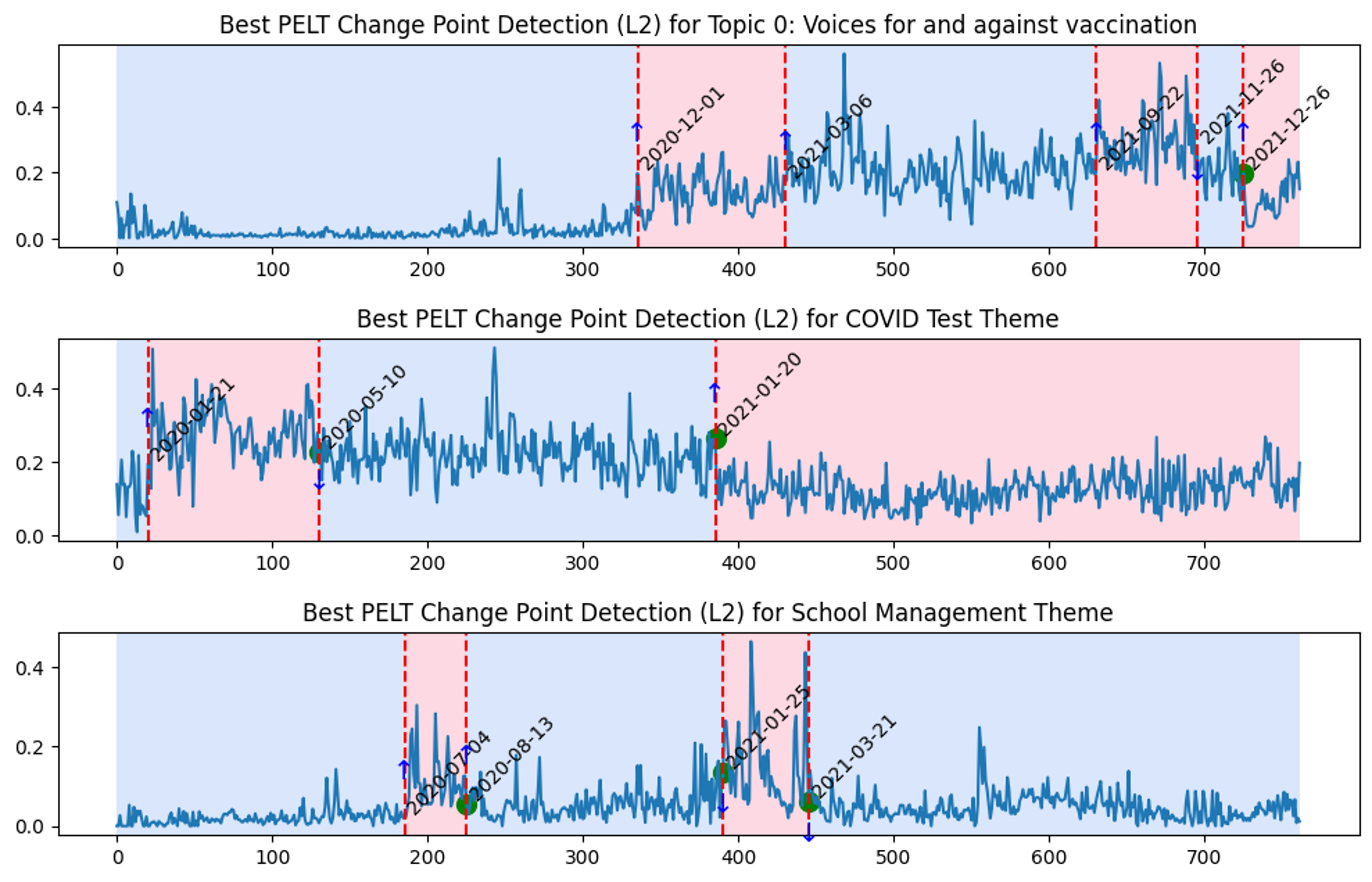}
    \caption{Later-phase dynamics of key CDC-related discussions detected by the PELT algorithm. Panels show (top) Topic 0 (pro- and anti-vaccination debates), (middle) COVID-19 testing discourse, and (bottom) school-management debates. Pink background highlights windows of heightened structural change, when discourse shifted in frequency or sentiment. Green dots indicate significant change-points, and blue arrows mark the direction of change, revealing how new variants and evolving guidelines repeatedly altered the trajectory of these conversations}
    \label{fig:PELT_themes}
\end{figure*}

During this acute phase, another prominent topic of discussion was the reliability and availability of COVID-19 testing (see Figure \ref{fig:PELT_themes}). Some users, less alarmed by the virus, argued that tests were overestimating case numbers, while others presented evidence suggesting under-reporting. A central debate revolved around whether the CDC was measuring active infections (via viral load) or past exposure (via antibodies) which would  inflate COVID numbers overall. Uncertainty about the accuracy of tests continued to be an issue of discussion in the following phases. 

\subsection{Middle Phase: Lock-downs and Masks}
The middle phase of the pandemic was marked by the introduction of lock-downs and the escalation of masking guidelines. After the second change-point for Topic 1 in late March 2020, discourse shifted toward Topic 30 (in-person gathering guidelines). Following the first U.S. lock-down on March 15\textsuperscript{th}, public conversation centered on risk management rather than questioning the legitimacy of COVID-19.

\begin{figure*}
    \centering
    \includegraphics[width=\textwidth]{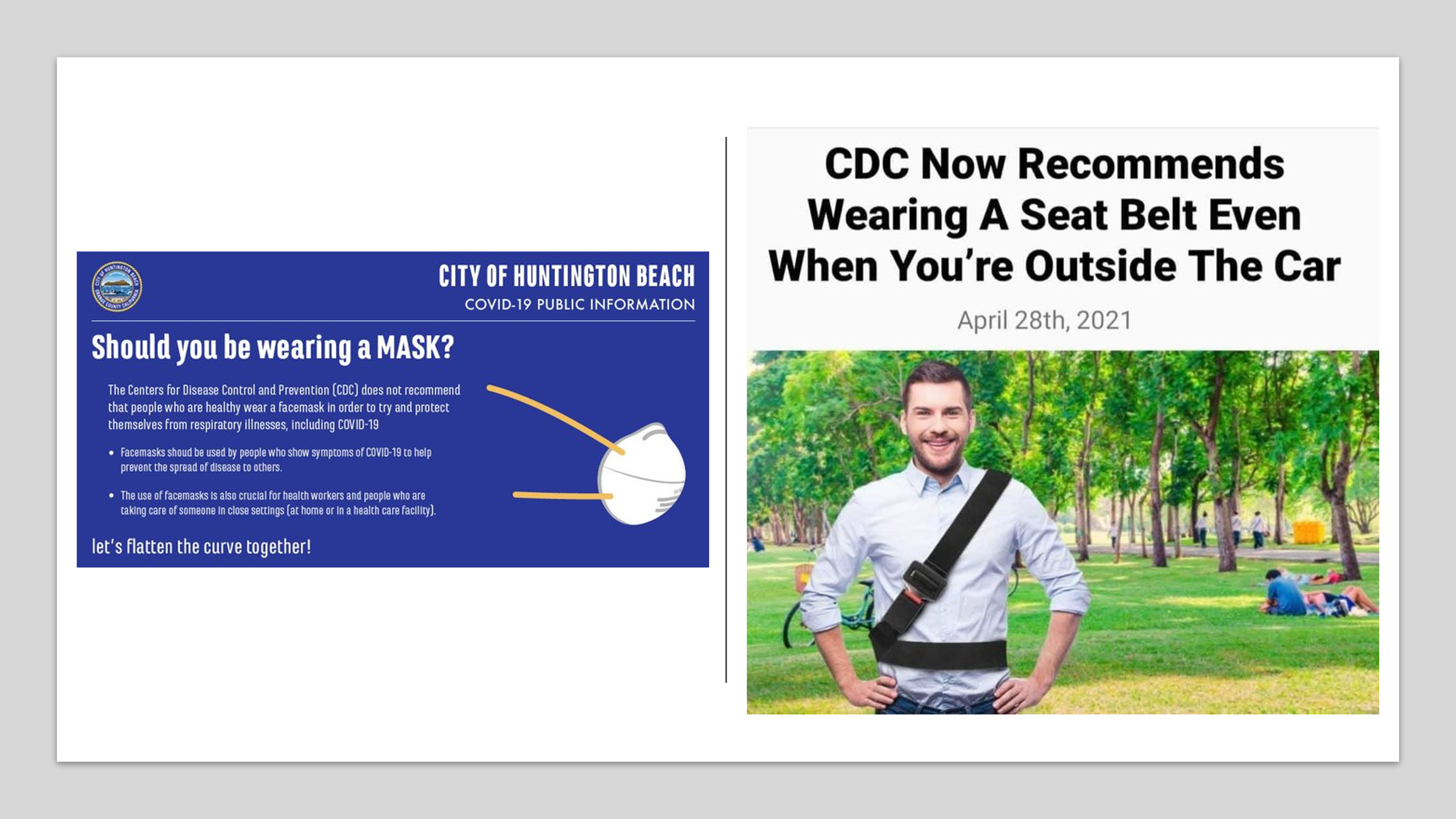}
    \caption{Contrasting user-shared ‘receipts’ highlighting shifting CDC mask guidance. Left: official local public-information graphic on when to wear a mask. Right: meme likening outdoor masking to wearing a seat belt outside a car, mocking what users viewed as excessive or inconsistent guidance}
    \label{fig:mask_diff}
\end{figure*}

Early discourse often critiqued what was seen as a lack of clear direction from the CDC, frequently through satirical takes on its guidelines. Many users characterized CDC recommendations—particularly around masking—as overly strict or inconsistent. Conversations about masking intensified between March 31\textsuperscript{st}, 2020, and June 14\textsuperscript{th}, 2021 (see Figure \ref{fig:PELT_topics}), as the CDC shifted from advising masks only for healthcare workers and symptomatic individuals to endorsing universal indoor masking. Figure \ref{fig:mask_diff} features a user-shared “receipt” that contrasted early and later CDC messages, mocking the perceived escalation in guidance. The post specifically referenced original guidelines, which restricted masking to medical personnel and those showing symptoms, to critique the CDC’s expanded recommendations for both indoor and outdoor settings. 

\begin{figure*}
    \centering
    \includegraphics[width=\textwidth]{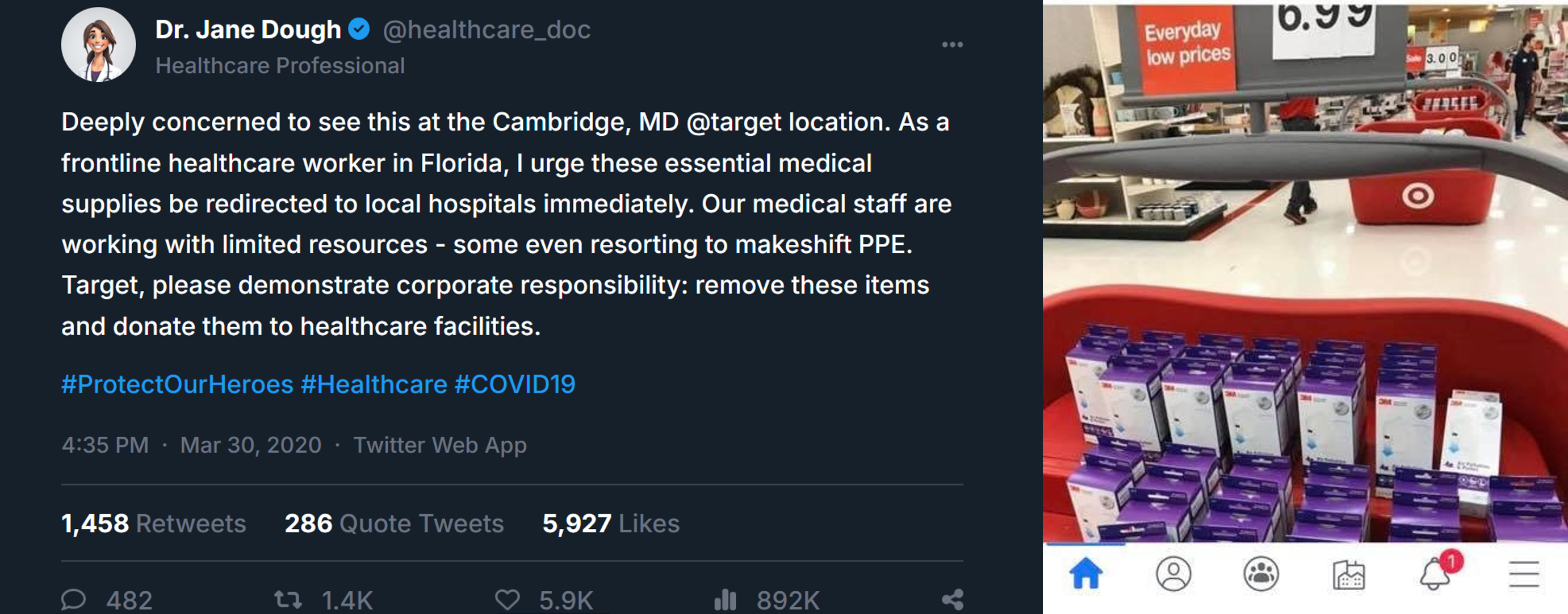}
    \caption{Example of an anonymized tweet circulated as evidence of early-pandemic PPE shortages in hospitals. Such posts were shared to contextualize or defend initial CDC guidance limiting mask use to healthcare workers, highlighting how frontline constraints influenced public discourse}
    \label{fig:doc_ref}
\end{figure*}

Many who criticized the evolving CDC guidelines—particularly the expanded masking recommendations—did so without recognizing the context behind earlier guidance. Specifically, they overlooked the initial scarcity of personal protective equipment (PPE), which heavily influenced the CDC’s early recommendations to reserve masks for healthcare workers and symptomatic individuals. As later guidelines promoted broader masking due to new scientific understanding of airborne transmission and asymptomatic spread, critics framed these changes as inconsistencies. However, in doing so, they failed to account for the earlier PPE shortages that had shaped the CDC’s original messaging. Some users, in contrast, shared “receipts” to remind others of these shortages in medical settings, as illustrated in Figure \ref{fig:doc_ref}.

This context, however, was not always remembered or acknowledged in public discourse. As a result, some questioned whether the CDC's guideline changes stemmed from scientific developments or were merely a strategic effort to redirect limited PPE to healthcare providers. This skepticism gave rise to hashtags like \#FauciDossier, which accused CDC officials—particularly Dr. Fauci—of deception dating back to the early pandemic. Although the “dossier” drew from low-credibility sources and included misinformation, some of its claims—such as shifts in masking guidance—were technically accurate, albeit presented in misleading and accusatory ways. This narrative of distrust continued to shape discourse into the pandemic’s later phases.

\subsection{End Phase: Post-Vaccine Variants and Evolving Guidelines} \label{subsec:end_phase}
\begin{figure*}
    \centering
    \includegraphics[width=\textwidth]{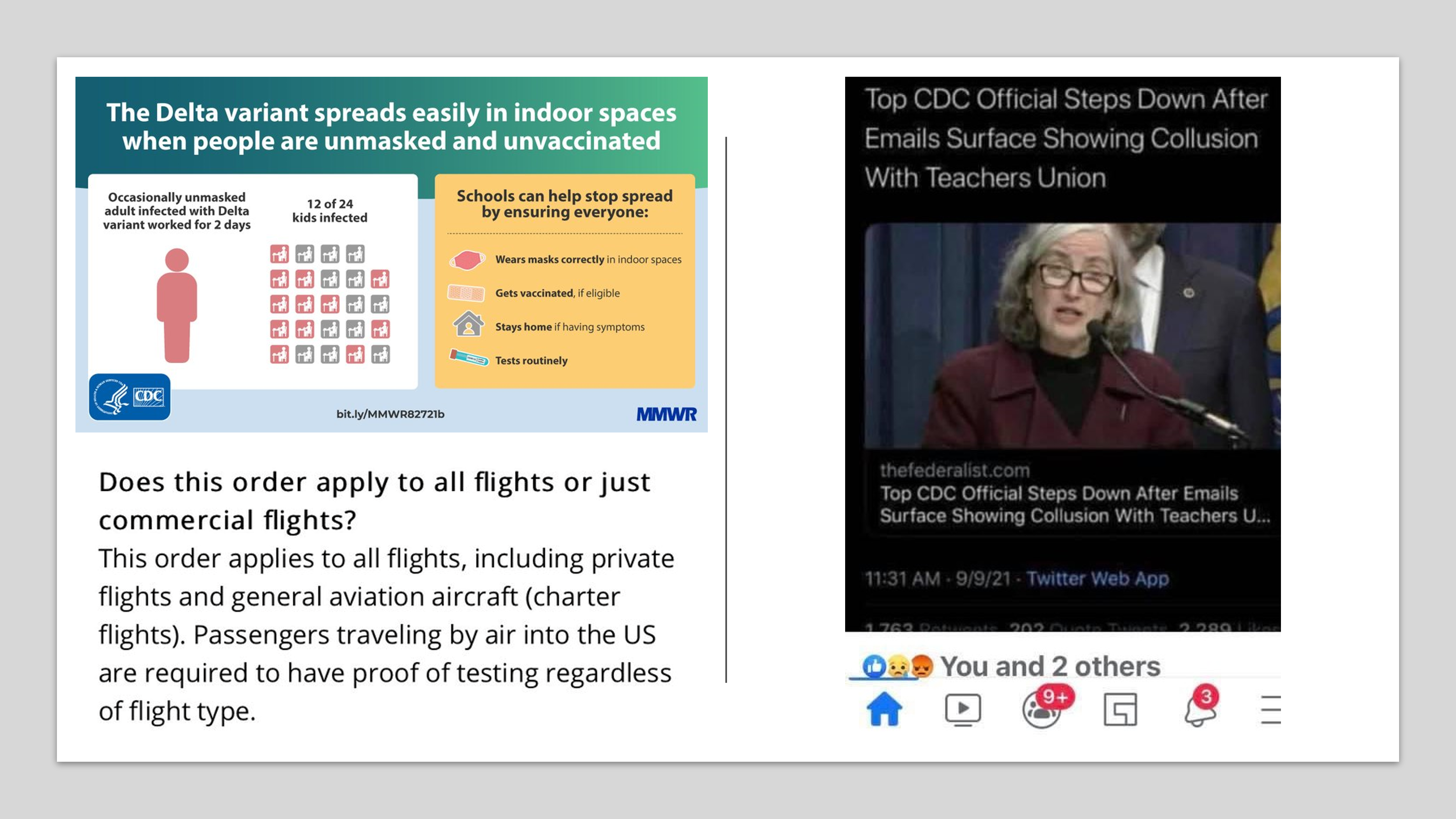}
    \caption{Images illustrating discourse around Delta-era school policies. Left: updated CDC guidance recommending universal masking in schools regardless of vaccination status. Right: a widely shared article speculating—without evidence—that teacher-union influence explained the guideline change.}
    \label{fig:change_perturbation}
\end{figure*}

In the final phase, public discourse was increasingly shaped by the rollout of vaccines and the emergence of new COVID-19 variants. While prior research has focused on vaccine misinformation and its role in fostering hesitancy \citep{bautista2024correcting}, my attention centers on the evolution of CDC messaging around vaccination—and how the public interpreted these shifts. Initially, CDC guidance indicated that fully vaccinated individuals no longer needed to wear masks or maintain social distancing (see Figure \ref{cdc_masking_change}). One notable example of this guidance was the lifting of mask requirements on airplanes for vaccinated travelers (see Figure \ref{fig:change_perturbation}). However, the rise of more transmissible variants complicated this narrative. As the virus mutated, public health recommendations adapted accordingly, resulting in messaging reversals that unsettled earlier assurances and fueled confusion over the ongoing protective value of vaccination.

One of the key change-points emerged within the School Management Theme. In the early stages of the pandemic, emphasis was placed on reopening schools with strict health protocols. Many parents believed that the introduction of vaccines would eventually make such measures unnecessary. However, as new variants—particularly the Delta variant—began to undermine vaccine effectiveness, updated CDC guidelines emphasized the continued importance of testing and masking alongside vaccination. These evolving recommendations sparked accusations of collusion between the CDC and teachers' unions (see Figure \ref{fig:change_perturbation}). Although unfounded, these claims reflected a broader misunderstanding of the rationale behind the shifting guidelines. In the absence of clear communication, many users relied on familiar frameworks to interpret the changing messages.

\begin{figure}[h!]
    \centering
    \includegraphics[width=1\linewidth]{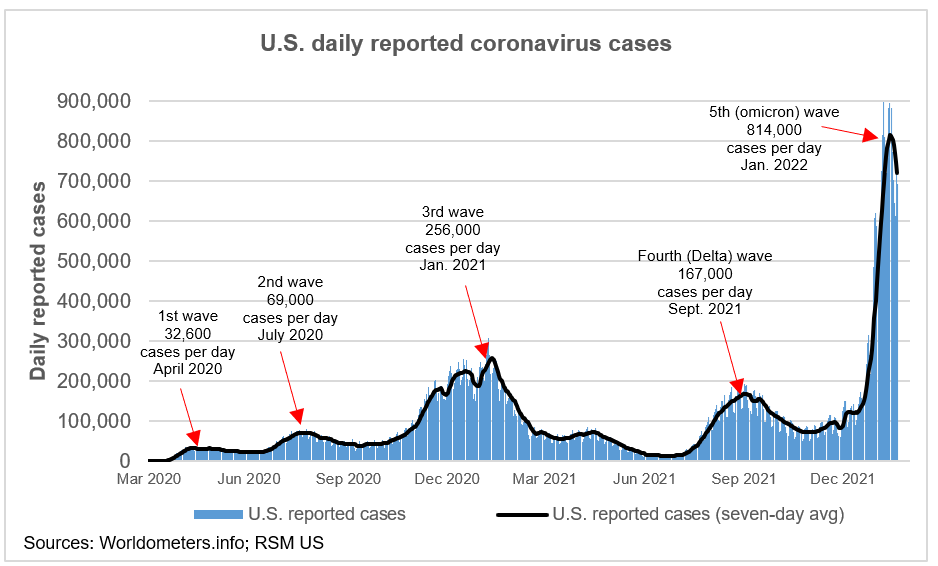}
    \caption{Timeline of U.S. daily reported COVID-19 cases across five major waves (March 2020–January 2022), highlighting the fourth (Delta) and fifth (Omicron) surges that triggered renewed debates about masking, vaccination efficacy, and testing.}
    \label{fig:omicron_delta}
\end{figure}

Another significant change-point was identified on December 26\textsuperscript{th}, 2021, aligning with the onset of the Omicron wave (Figure \ref{fig:omicron_delta}). Topic 0, which captured voices both supporting and opposing vaccination, saw a notable surge, along with Topics 47 and 31, which addressed how the Omicron and Delta variants affected masking practices and vaccine efficacy. These discussions reflected growing concern that emerging variants were reducing the effectiveness of existing vaccines, prompting renewed calls for mitigation strategies such as masking and distancing (Topics 30, 38, and 62). A further point of contention during this period was the perceived decline in testing accuracy. As newer variants spread, debates intensified over the reliability of COVID-19 tests, especially as less-accurate results began to undermine confidence in public health efforts.

\begin{figure*}
    \centering
    \includegraphics[width=\textwidth]{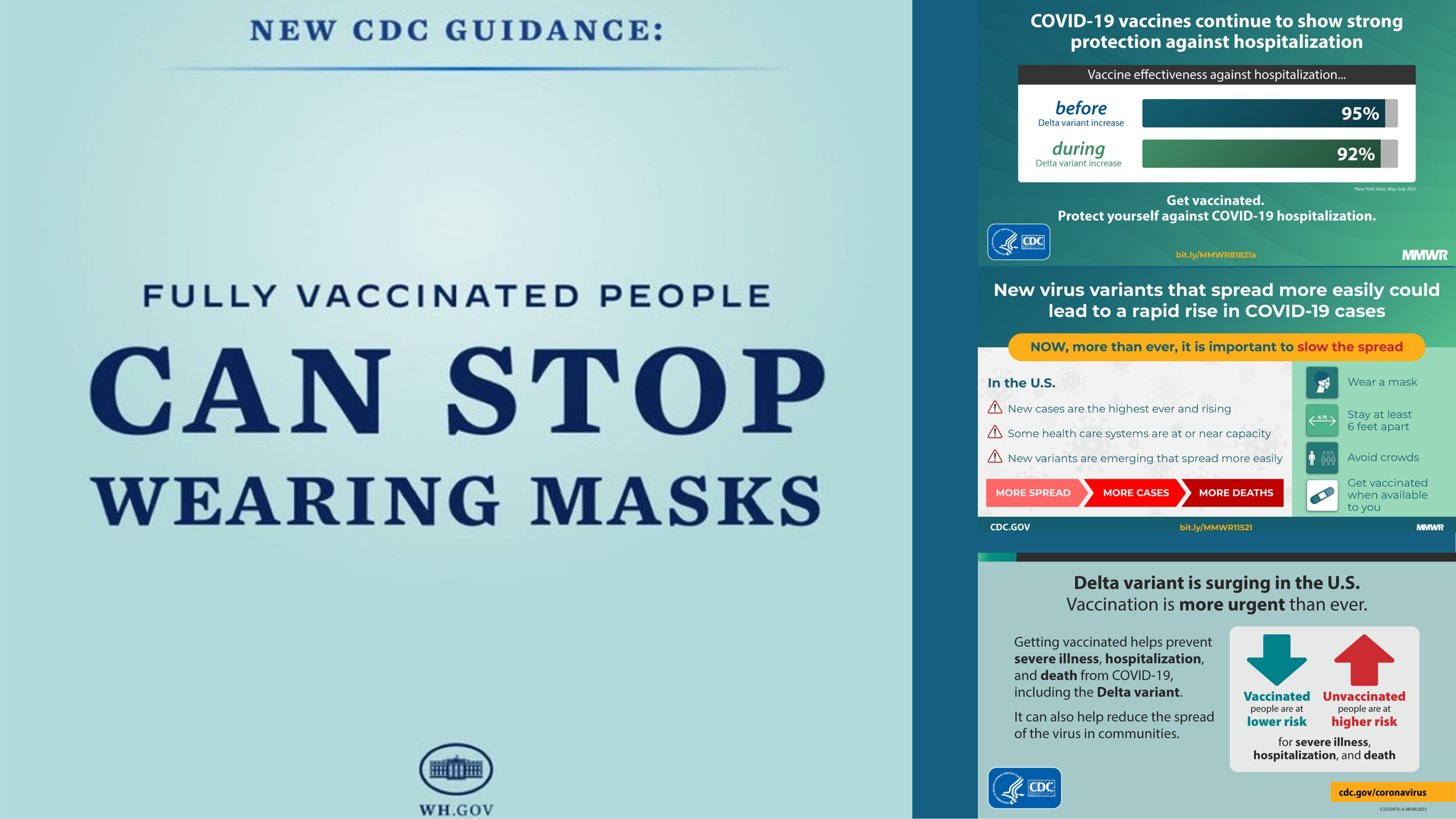}
    \caption{Contradictory perceptions of CDC vaccine guidance. Left: a prominent ‘receipt’ of an early CDC message suggesting fully vaccinated people could stop masking. Right: later CDC updates clarifying that vaccination reduces severe disease but—given new variants—masking and other measures may still be required.}
    \label{fig:cdc_masking_change}
\end{figure*}

A large number of users circulated a “receipt” featuring a CDC guideline, amplified by the White House, stating that fully vaccinated individuals no longer needed to wear masks (see Figure \ref{fig:cdc_masking_change}). They argued that subsequent guidelines contradicted this statement, which was presented without any explicit caveats. Although the CDC later issued several nuanced updates—three of which are shown on the right side of Figure \ref{fig:cdc_masking_change}—these clarifications did little to quell confusion. Many users questioned how new variants could seemingly “appear out of nowhere” and so quickly alter the protective value of vaccination. This skepticism gave rise to conspiracy theories, including claims that the constant push for updated vaccines was driven by pharmaceutical companies seeking sustained profits.

\subsection{Change-points in CDC Tweet Frequency, Frequency of Rich Media Content, and Overall Sentiment}

\label{sec:evolve_over_time_2}

Examining media use, I found two significant change-points in the frequency of Tweets containing rich media: February 25\textsuperscript{th}, 2020, and December 4\textsuperscript{th}, 2020. Both exhibited upward trends. After the first change-point, Tweets increasingly included images conveying detailed information on mitigation guidelines, particularly focused on topics such as social distancing guidelines (Topic 62), facial hair and masking guidelines (Topic 57), and controversies surrounding Hydroxychloroquine and Ivermectin (Topic 48).

Following the second change-point, rich media usage shifted toward emphasizing updates related to the Delta and Omicron variants, which altered masking guidelines, particularly around Topics 4, 31, and 47. This period coincided with renewed calls for masking, as the new variants were seen to diminish vaccine efficacy. Additionally, many Tweets during this time incorporated "receipts"—screenshots of earlier CDC statements—to critique the evolving guidelines.

\begin{figure*}
    \centering
    \includegraphics[width=\textwidth]{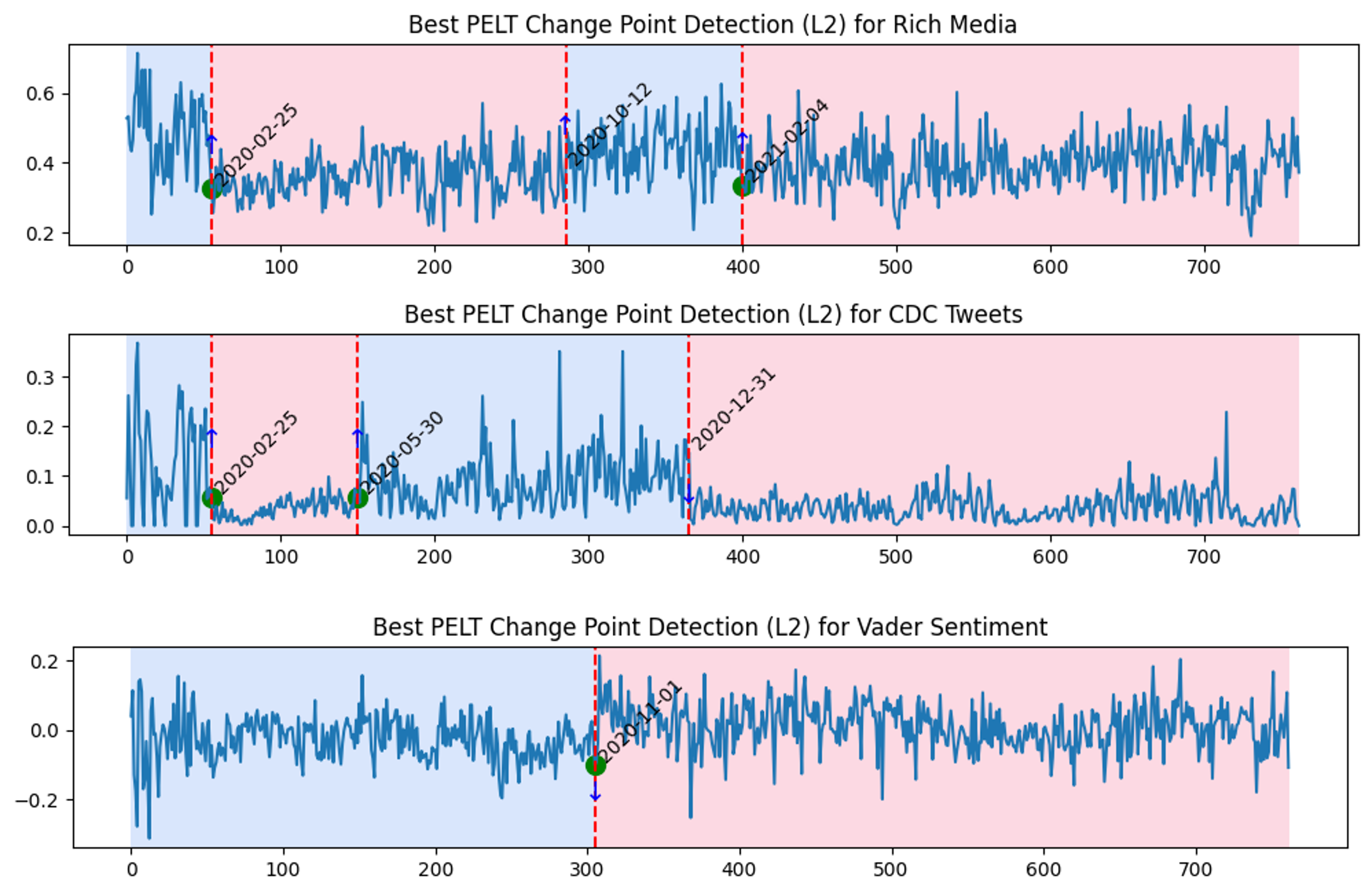}
    \caption{Temporal change-points in platform activity and sentiment revealed by the PELT algorithm. Panels show (top) frequency of rich-media use in CDC-related tweets, (middle) tweet volume from the CDC’s official account, and (bottom) overall sentiment measured by VADER. Pink-shaded areas indicate intervals of intensified structural change, when tweet composition or tone shifted substantially. Green dots denote significant change-points and blue arrows show whether activity or sentiment trended upward or downward across these intervals, capturing the evolving rhythm of CDC messaging and public response}
    \label{fig:PELT_photo_verified}
\end{figure*}

The PELT time series analysis, shown in Figure \ref{fig:PELT_photo_verified}, reveals that CDC Tweet volume dropped significantly between February 25\textsuperscript{th}, 2020, and late May 2020. During this vacuum, public discourse increasingly critiqued the CDC, especially through parodies of the CDC's masking guidance (Topic 59) and debates around the Hydroxychloroquine and Ivermectin controversy (Topic 48).

Sentiment trends, as measured by VADER scores, also shifted over time. Sentiment was initially positive until a change-point on November 1\textsuperscript{st}, 2020, after which it became significantly negative, largely driven by discourse around political and legislative responses to COVID-19 (Topic 29). Later stages of discussion revealed growing public frustration with renewed restrictions during the holiday season and with mitigation policies responding to new variant waves.

\subsection{User Commitment Over Time}
\label{sec:survival_results}

The Cox Proportional Hazards model yielded a concordance of 0.84, indicating high predictive performance in modeling user persistence in CDC-related discourse. The model identifies predictors of longer or shorter engagement durations, grouped by feature type: platform affordances, linguistic and sentiment features, and topic-level discourse.

\subsubsection{Platform Affordances}

Verified users were more likely to remain active (\textit{HR} = 0.89, $p < 0.0001$), as were users whose tweets received more retweets (\textit{HR} = 0.93, $p < 0.05$). Sharing links to low-credibility sources also predicted sustained engagement (\textit{HR} = 0.94, $p < 0.0001$), possibly due to their oppositional tone. Being listed in public Twitter lists was also mildly protective (\textit{HR} = 0.99, $p < 0.05$).

In contrast, higher like counts predicted shorter participation (\textit{HR} = 1.08, $p < 0.05$), indicating that passive affirmation may reduce discourse continuity. No platform affordances significantly increased engagement risk beyond these factors.

\begin{table*}
    \centering
    \caption{Shows how features of the Twitter platform—such as verified status, retweet counts, low-credibility source use, list membership, and like counts—statistically influenced how long users remained active in CDC-related discussions. Lower hazard ratios (HR) indicate factors associated with longer engagement, while HR > 1 indicates quicker disengagement}
    \label{tab:platform-affordances}
    \begin{tabular}{|l|c|c|}
        \hline
        \textbf{Feature} & \textbf{Hazard Ratio (HR)} & \textbf{p-value} \\
        \hline
        Verified User & 0.89 & $<0.0001$ \\
        Retweet Count & 0.93 & $<0.05$ \\
        Low-Credibility Source Use & 0.94 & $<0.0001$ \\
        Listed Count & 0.99 & $<0.05$ \\
        Like Count & 1.08 & $<0.05$ \\
        \hline
    \end{tabular}
\end{table*}

\subsubsection{Linguistic and Sentiment Features}

The strongest predictor of prolonged engagement was cognitive process language (\textit{HR} = 0.69, $p < 0.0001$), followed by insight (\textit{HR} = 0.86, $p < 0.0001$), power-related drives (\textit{HR} = 0.85, $p < 0.0001$), and achievement-related terms (\textit{HR} = 0.94, $p < 0.0001$). Social references (\textit{HR} = 0.92, $p < 0.0001$), motion verbs (\textit{HR} = 0.93, $p < 0.0001$), and expressions of tentativeness (\textit{HR} = 0.91, $p < 0.0001$) also correlated with sustained discourse.

Temporal orientation mattered: users referencing the past (\textit{HR} = 0.92, $p < 0.0001$) or future (\textit{HR} = 0.96, $p < 0.0001$) remained engaged longer. Other predictors of longevity included negations (\textit{HR} = 0.96), informal language (\textit{HR} = 0.95), cause (\textit{HR} = 0.87), comparison (\textit{HR} = 0.96), quantifiers (\textit{HR} = 0.96), perception (\textit{HR} = 0.96), and first-person singular pronouns (\textit{HR} = 0.97)—all significant at $p < 0.0001$.

Emotionally, negative affect predicted retention: sadness (\textit{HR} = 0.98), anger (\textit{HR} = 0.98), anxiety (\textit{HR} = 0.99), and sexuality-related terms (\textit{HR} = 0.98) all lowered hazard. Meanwhile, general affect words (\textit{HR} = 1.05, $p < 0.0001$) and positive sentiment (VADER: \textit{HR} = 1.06, $p < 0.0001$) were associated with shorter engagement.

Syntactic complexity predicted dropout: high use of verbs (\textit{HR} = 1.42), prepositions (\textit{HR} = 1.12), and conjunctions (\textit{HR} = 1.04) were all linked to shorter tenure. Linguistic informality such as netspeak (\textit{HR} = 1.02, $p < 0.01$) and nonfluencies (\textit{HR} = 1.01, $p < 0.01$) also correlated with faster disengagement.

\begin{table*}
    \centering
    \caption{Summarizes the key linguistic styles and emotional expressions that shaped the longevity of participation. Cognitive process words, insight and power-related drives, temporal focus, and negative affect all promoted sustained dialogue, whereas high syntactic complexity and positive sentiment predicted shorter involvement}
    \label{tab:linguistic-sentiment}
    \begin{tabular}{|l|c|c|}
        \hline
        \textbf{Feature} & \textbf{Hazard Ratio (HR)} & \textbf{p-value} \\
        \hline
        Cognitive Process Words & 0.69 & $<0.0001$ \\
        Drives: Power & 0.85 & $<0.0001$ \\
        Insight Words & 0.86 & $<0.0001$ \\
        Discrepancy Words & 0.89 & $<0.0001$ \\
        Certain Words & 0.90 & $<0.0001$ \\
        Tentativeness & 0.91 & $<0.0001$ \\
        Focus on Present & 0.93 & $<0.0001$ \\
        Focus on Past & 0.92 & $<0.0001$ \\
        Focus on Future & 0.96 & $<0.0001$ \\
        Motion Words & 0.93 & $<0.0001$ \\
        Social Words & 0.92 & $<0.0001$ \\
        Cause Words & 0.87 & $<0.0001$ \\
        Negations & 0.96 & $<0.0001$ \\
        Compare Words & 0.96 & $<0.0001$ \\
        Quantifiers & 0.96 & $<0.0001$ \\
        Perception Words & 0.96 & $<0.0001$ \\
        Sexuality Words & 0.98 & $<0.0001$ \\
        Sadness Words & 0.98 & $<0.0001$ \\
        Anger Words & 0.98 & $<0.0001$ \\
        Anxiety Words & 0.99 & $<0.01$ \\
        “I” Pronoun (1st Person Singular) & 0.97 & $<0.0001$ \\
        Informal Language & 0.95 & $<0.0001$ \\
        Verb Usage & 1.42 & $<0.0001$ \\
        Prepositions & 1.12 & $<0.0001$ \\
        Conjunctions & 1.04 & $<0.0001$ \\
        Affect Words (General) & 1.05 & $<0.0001$ \\
        VADER Sentiment Score (Positive) & 1.06 & $<0.0001$ \\
        Netspeak & 1.02 & $<0.01$ \\
        Nonfluencies & 1.01 & $<0.01$ \\
        \hline
    \end{tabular}
\end{table*}

\subsubsection{Topic-Level Engagement}

Topic engagement had a more modest but interpretable effect. Users who engaged with the “Questioning Institutions” theme were more likely to stay involved (\textit{HR} = 0.95, $p < 0.001$). In contrast, users focused on rapidly evolving but institutionally framed themes—including vaccination debates (Topic 0: \textit{HR} = 1.02), legal/political commentary (Topic 29: \textit{HR} = 1.02), gathering guidelines (Topic 30: \textit{HR} = 1.02), variant updates (Topic 31: \textit{HR} = 1.01), school policies (Topic 38: \textit{HR} = 1.01), and social distancing (Topic 62: \textit{HR} = 1.01)—were all significantly more likely to disengage quickly ($p < 0.0001$).

\begin{table*}
    \centering
    \caption{Details how engagement with specific discourse topics predicted persistence. Skeptical “Questioning Institutions” discussion extended participation, while conversations about vaccines/variants, political framing, and specific mitigation guidelines were linked to earlier dropout}
    \label{tab:topic-engagement}
    \begin{tabular}{|l|c|c|}
        \hline
        \textbf{Topic} & \textbf{Hazard Ratio (HR)} & \textbf{p-value} \\
        \hline
        Questioning Institutions & 0.95 & $<0.001$ \\
        Topic 0: Vaccine/Variant Debates & 1.02 & $<0.0001$ \\
        Topic 29: Political/Legal Framing & 1.02 & $<0.0001$ \\
        Topic 30: In-Person Gathering Guidelines & 1.02 & $<0.0001$ \\
        Topic 31: Delta/Omicron Updates & 1.01 & $<0.0001$ \\
        Topic 38: School Policies & 1.01 & $<0.0001$ \\
        Topic 62: Social Distancing Guidelines & 1.01 & $<0.0001$ \\
        \hline
    \end{tabular}
\end{table*}

\subsubsection{Overall Summary}

\paragraph{Breakpoint Summary}
Parallel change-point detection using the PELT algorithm highlighted key structural disruptions in discourse and media use. Breakpoints aligned with major pandemic milestones: early 2020 hoax debates; spring 2020 lockdown and mask-guidance shifts; late 2020 negativity around politics; and late 2021 Delta/Omicron variant surges. These inflection points captured abrupt changes in tweet volume, sentiment, and rich-media usage (e.g., spikes in “receipt” screenshots), signaling moments when public conversation—and trust—pivoted sharply,

\paragraph{Survival Analysis Summary}
The Cox proportional hazards model (concordance = 0.84) reveals that long-term engagement was not driven by positive affirmation but by critical, cognitively complex discourse. Verified users and those frequently retweeted or sharing low-credibility links tended to remain active. Linguistic signals such as cognitive process words, expressions of insight, power-related language, and temporally oriented speech (past and future references) were strongly protective against disengagement. Negative emotions like sadness and anger also contributed to retention, while positive affect and syntactic density accelerated exit. At the topical level, “Questioning Institutions” fostered persistence, whereas engagement focused on concrete CDC updates or policy debates predicted shorter tenure.

\section{Discussion} \label{subsec:disc_rq3}
In the discussion, we synthesize these empirical findings to show how \emph{CDC guidance became an evolving ``crisis messaging journey.''}

We first interpret the \emph{communication gaps and sustained public concern} revealed by our data, arguing that persistent skepticism and politically inflected distrust prolonged engagement well beyond the acute phases of the pandemic.

We then elaborate the concept of \emph{crisis messaging journeys} to explain how CDC statements were continually repurposed as ``receipts,'' generating data friction and inclusion friction that complicated efforts to update guidance in light of new evidence.

Building on these insights, we propose \emph{design recommendations}---from embedding explicit caveats and annotated quote-tweets to pre-bunking and real-time flashpoint detection---that could help public health agencies communicate more transparently and adaptively across prolonged crises.

We conclude by reflecting on \emph{limitations and future research}, including the need to examine visual framing and to pair computational methods with qualitative interviews, and by outlining how our framework extends crisis informatics and CSCW scholarship toward the design of more resilient public health communication infrastructures.

\subsection{When the Crisis Doesn’t End: Communication Gaps and Ongoing Public Concern}
Most studies of crisis informatics focus on the response of crisis organization (e.g., the CDC) during early phases of a crisis \cite{soden_palen_2018}. However, even in early 2025, on the eve of the fifth anniversary of COVID, around 40\% of the population still think the pandemic is ``not over,'' that they are not back to living their ``normal'' lives, and are worried about the long-lasting effects of COVID (e.g., long COVID) and of the next health crises (e.g., bird flu) \cite{gallup2025pandemic}. This shows that there is a need to understand the long-term communication response needs for public health crises like COVID. 

The findings of this study reveal that questioning the CDC's communication strategies played a critical role in sustaining user engagement over time. Accounts that expressed skepticism or distrust toward institutional responses—particularly those highlighting political interference, inconsistent messaging, or perceived inefficiencies—demonstrated greater longevity within the CDC public sphere. This pattern echoes historical challenges in public health communication. During the 1894 smallpox outbreak in Milwaukee, Health Commissioner Walter Kempster remarked, "But for politics and bad beer, the matter would never have been heard of" \citep{leavitt2003public}[P.187-188], signaling how political and communicative failures contributed to the erosion of civic trust. Despite Kempster’s medical expertise, he overlooked the essential role of public dialogue and responsive communication in managing health crises. Similarly, the endurance of critical voices during the COVID-19 pandemic underscores how distrust toward public institutions—exacerbated by communication breakdowns—can sustain public discourse and prolong engagement in times of crisis.

\subsection{Crisis Messaging Journeys: Data Friction and the Evolving Life of Public Health Guidance}
While earlier cases illustrate how crucial effective communication is to managing health crises, contemporary challenges underscore how difficult this remains. Most studies of crisis informatics have focused on organizational responses during the early phases of a crisis (e.g., the CDC's initial efforts during COVID-19; \cite{soden_palen_2018}). However, even in early 2025—five years after the pandemic began—around 40\% of Americans continued to believe that COVID-19 was "not over," felt they had not returned to "normal" life, and remained concerned about long-term impacts such as long COVID and future threats like avian influenza \citep{gallup2025pandemic}. This enduring sense of crisis highlights the urgent need to understand and design for the long-term communication and engagement demands of extended public health emergencies.

Dalrymple \cite{dalrymple2016facts} argues that there is a need for a rapid change of communication strategy around these flashpoints, and that, due to social media affordances, especially persistence and forwardability, changing communication strategies can be difficult as responses to the CDC included ``receipts'' of earlier CDC messaging. My findings show that this reaction to CDC communications might be explained by the perceived lack of honesty about the CDC's uncertainty vis-a-vis an emerging infectious disease \citep{holmes2008communicating}. In fact, Holmes et al. \cite{j2009communicating} suggested that focusing on ``certainty for now'' could earn public trust more than conveying certainty and then being proven wrong about IEDs in future flashpoints.

This is in line with earlier work that shows how data friction (challenges of sharing and interpreting data; \cite{edwards2011science}) increases with the distance between message creators and re-users \citep{borgman2025data} and with the duration of time from the original message to the time of reuse \citep{bates2016data,borgman2025data}. The different ``ways in which [data] are ...modified to fit different uses as they travel across space, time and social situations'' are defined as data journeys \citep{leonelli2020learning}. I introduce the concept of \textit{crisis messaging journeys} which describes how crisis messages initially shared by the CDC change based on context, purpose (e.g., opposing new mitigation guidelines), and time elapsing from the initial CDC message \citep{gregory2024data}. 

Guidelines shared by the CDC at any time during the pandemic might change later given new information gleaned about the EID (e.g., while the vaccine is still effective against hospitalization, people can still be infected with new variants). While scientists understand that ``scientific knowledge is always provisional, always open to revision,'' \citep[P.427]{edwards2013vast} public engagement with scientific knowledge (i.e., members of the public using ``receipts'' of earlier guidelines to argue against the veracity of updated guidelines) can ``represent ideological and political strategies [that] represent a new form of friction...inclusion friction [which can] promote confusion, suspicion,'' and misinformation \cite[ibid]{edwards2013vast}. 

\subsection{Design Recommendations to Manage Flashpoints throughout Crisis Message Journeys}
There are three ways to reduce inclusion friction: (1) using more cautious language by employing caveats in public science communication \citep[P.200]{stocking1993constructing} when presenting guidelines signifying ``this is what we know now, rather than this is what we know''\citep{gregory2024data}; (2) linking the original message either by quote-tweeting it to explicitly show the evolution of science and/or annotating \citep{zheng2022evaluating} the original message using an affordance like community notes \citep{Kankham11092024} or redaction notices (described in \S\ref{related_work_1}); and (3) pre-exposure \citep[P.116]{lewandowsky2012misinformation} or pre-bunking \citep{barman2024evaluating} where the audience is presented with an alternative narrative \citep[P.117]{lewandowsky2012misinformation}, thus filling any gaps which might result from changing messages in a flashpoint. This would include messaging to explain the reasons for new guidelines by comparing them with their original messages and explaining how the difference was brought about by new medical information. 

Earlier work has shown that if uncertainty claims are made early by scientists in their public science communication, they are seen as being frank as they ``openly admit incomplete knowledge.'' \citep[P.9]{zehr2012scientists} This is imperative to the success of the crisis organization because the publics' audiences are more ``likely to reject organizational messages during a crisis situation that are inconsistent with past actions'' \citep[P.399]{keri_et_al_2005} which would negatively effect crisis communication around flashpoints.

Given the importance of identifying flashpoints and changing communication strategies accordingly \citep{dalrymple2016facts}, automating the detection of change-points can help the CDC deal with flashpoints as they occur. Based on a system presented by \cite{lyu2024bipec}, I would first apply a baseline change-point data set from earlier interactions which are then incorporated into a Bayesian model. This first phase provided one of two outputs: either the change is a candidate for a change-point or it is not. If it is, a PELT model is applied at set time durations (e.g., every week or month). When detecting a change-point, manual inspection is still important to verify that indeed there is a change after the flashpoint. The manual coder can confirm that indeed, this is a flashpoint, remove it, or if not clear, set it as pending to request annotation by another human-in-the loop. This will allow the CDC to change its communication strategy. 

\subsection{Limitations}
Given the power of visual media in public health communication, future research should apply critical discourse analysis to cluster and track changes in visual framing across the crisis timeline \cite{walters_et_al_24}. Additionally, although our study draws from a substantial dataset, I did not engage directly with users to explore their perceptions of COVID-19 discourse within the CDC arena. Building on \cite{Pine_et_al_21}'s work, future studies should incorporate interviews to better understand not only how users consumed healthcare information, but also how they contributed to the conversation or transitioned into new online communities.

\section{Conclusion}
Public health communication during protracted crises is an ongoing negotiation of knowledge. By linking survival modeling with change-point detection, this study shows how institutional uncertainty, critical publics, and viral “receipts” sustained debate beyond initial outbreaks. Rather than a one-way flow of facts, CDC messaging became a contested epistemic arena where flashpoints repeatedly reset discourse. Future crisis communication should combine cautious, annotated messaging with real-time flashpoint detection and human-in-the-loop oversight to foster trust and resilience over time.

\bibliographystyle{ACM-Reference-Format}
\bibliography{sample-base}

\end{document}